\documentclass[
    aps,
    prd,
    reprint,
    superscriptaddress,
    nofootinbib,
    longbibliography
]{revtex4-2}

\usepackage[T1]{fontenc}
\usepackage[utf8]{inputenc}
\usepackage{amsmath,amssymb,bm}
\usepackage{graphicx}
\usepackage{booktabs}
\usepackage{xcolor}
\usepackage{microtype}
\usepackage[
    colorlinks=true,
    linkcolor=blue,
    citecolor=blue,
    urlcolor=blue
]{hyperref}

\newcommand{\HaloIntegralFigurePath}{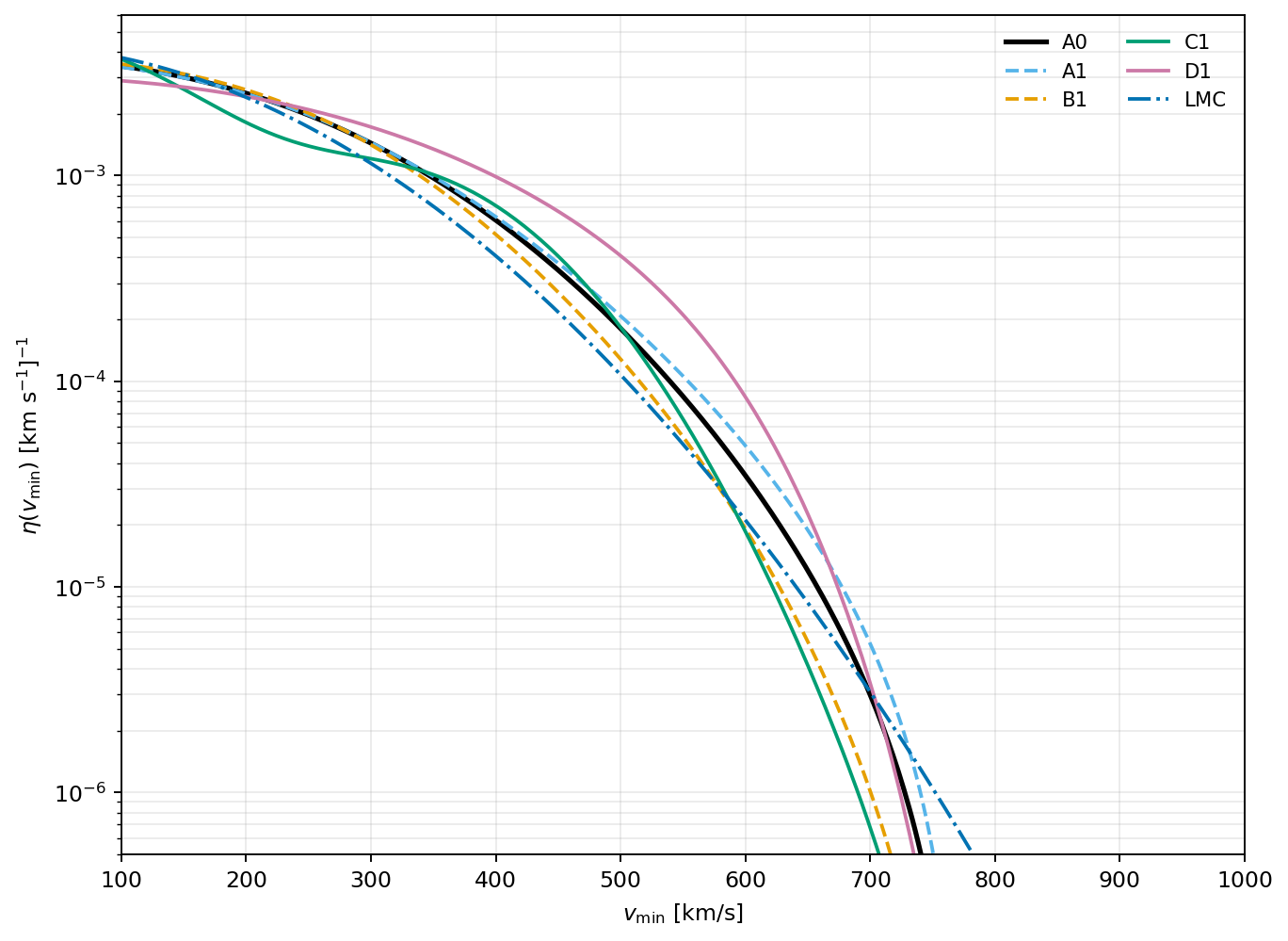}
\newcommand{\SIFigurePath}{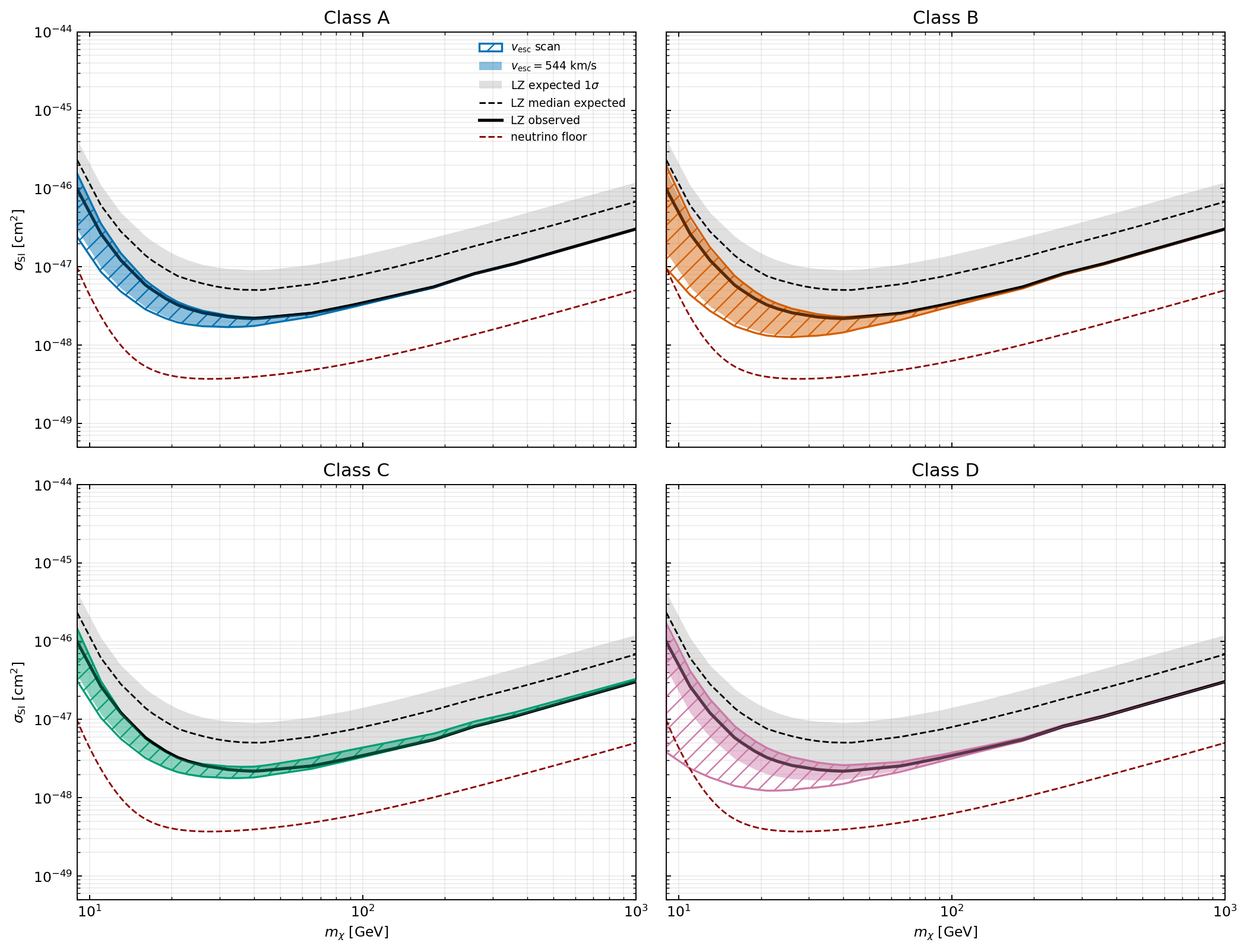}
\newcommand{\SDFigurePath}{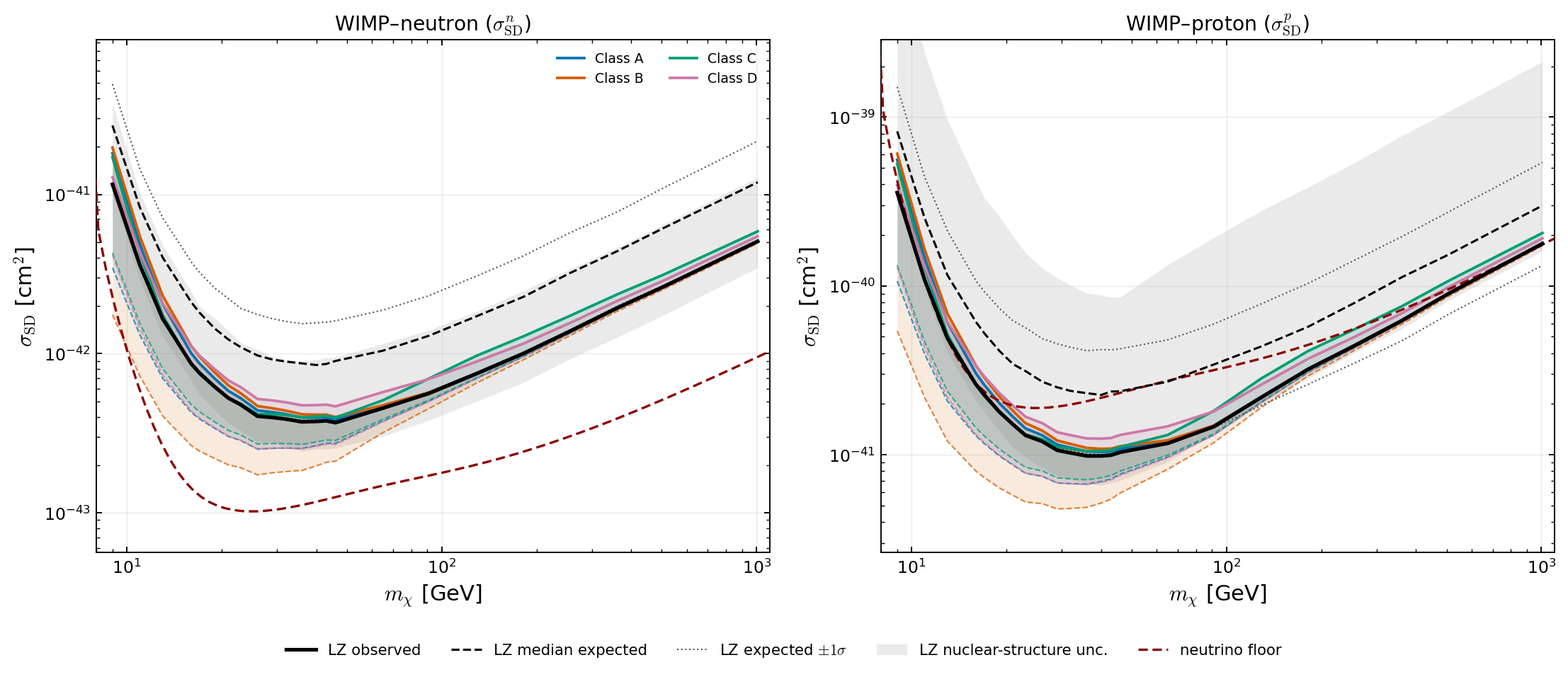}
\newcommand{\InelasticFigurePath}{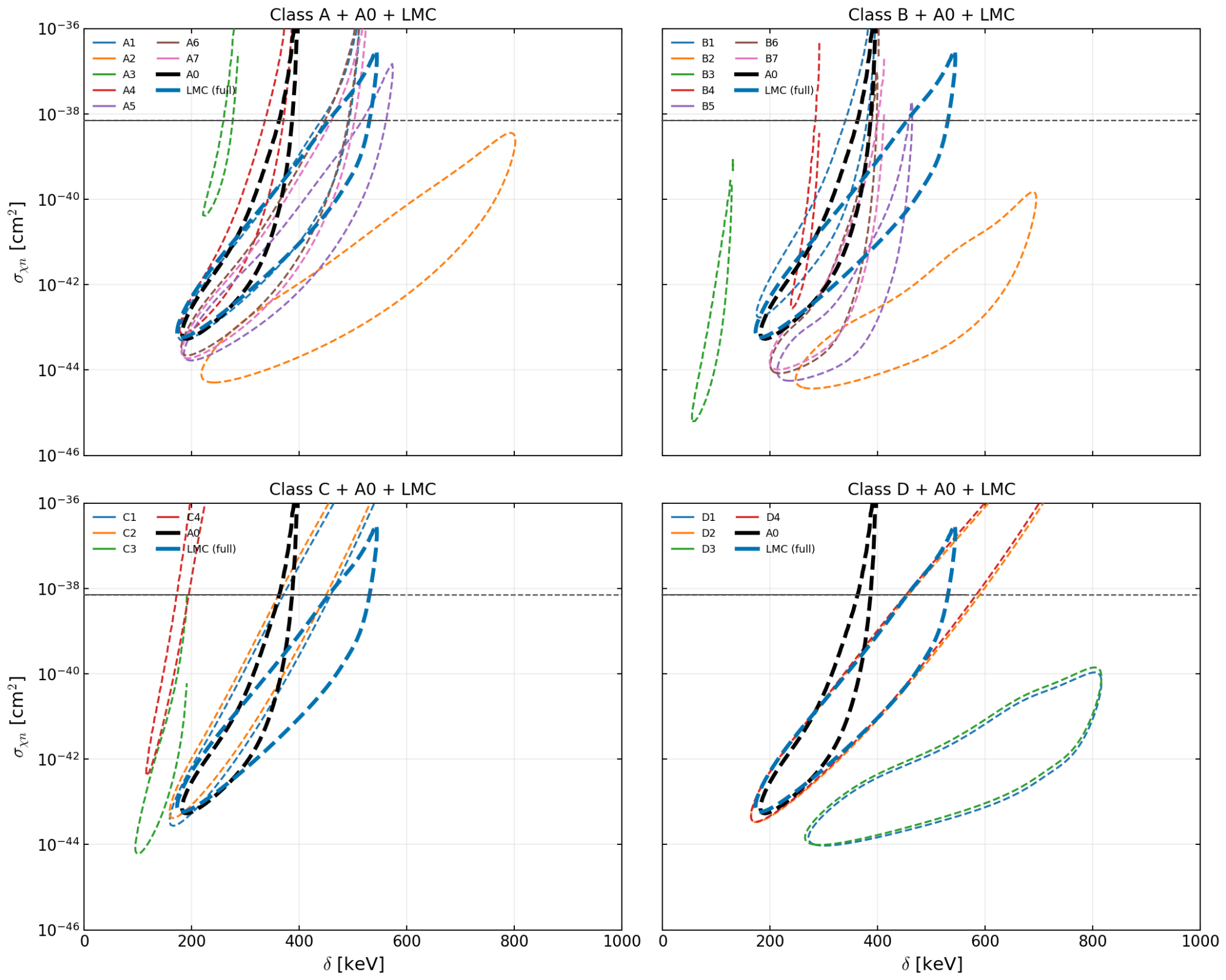}

\makeatletter
\newcommand{\SafeIncludeGraphics}[3][]{%
    \begingroup
    \edef\ResolvedFigurePath{#2}%
    \ifx\ResolvedFigurePath\@empty
        \fbox{%
            \parbox[c][0.20\textheight][c]{0.90\linewidth}{%
                \centering
                \textbf{Figure placeholder}\\[0.8em]
                #3\\[0.8em]
                \footnotesize
                Insert the original image path in the preamble.
            }%
        }%
    \else
        \IfFileExists{\ResolvedFigurePath}{%
            \includegraphics[#1]{\ResolvedFigurePath}%
        }{%
            \fbox{%
                \parbox[c][0.20\textheight][c]{0.90\linewidth}{%
                    \centering
                    \textbf{Image file not found}\\[0.8em]
                    #3\\[0.8em]
                    \footnotesize
                    Check the original path and uploaded file.
                }%
            }%
        }%
    \fi
    \endgroup
}
\makeatother

\newcommand{\dd}{\mathrm{d}}
\newcommand{\GeV}{\mathrm{GeV}}
\newcommand{\TeV}{\mathrm{TeV}}
\newcommand{\keV}{\mathrm{keV}}
\newcommand{\kpc}{\mathrm{kpc}}
\newcommand{\kms}{\mathrm{km\,s^{-1}}}
\newcommand{\cmSq}{\mathrm{cm^2}}

\newcommand{\vesc}{v_{\mathrm{esc}}}
\newcommand{\vmin}{v_{\min}}
\newcommand{\vmax}{v_{\max}}
\newcommand{\ER}{E_R}
\newcommand{\Eobs}{E_{\mathrm{obs}}}

\newcommand{\mchi}{m_\chi}
\newcommand{\rholoc}{\rho_0}

\newcommand{\sigman}{\sigma_{\chi n}}

\newcommand{\SHM}{\mathrm{SHM}}

\newcommand{\eff}{\epsilon_{\mathrm{eff}}}

\begin{document}

\title{
Quantifying the Impact of Galactic Dark Matter Halo Models
on LZ Results
}

\author{Jialiang Shao}
\affiliation{Department of Astronomy, School of Physical Sciences, University of Science and Technology of China, Hefei 230026, China}

\author{Andrea Addazi}
\email{addazi@aqnu.edu.cn}
\affiliation{School of Physics and Astronomy, Anqing Normal University, Anqing 246133, China}
\affiliation{Institute of Astronomy and Astrophysics, Anqing Normal University, Anqing 246133, China}
\affiliation{CAS Key Laboratory for Research in Galaxies and Cosmology, School of Astronomy and Space Science, University of Science and Technology of China, Hefei 230026, China}
\affiliation{Laboratori Nazionali di Frascati INFN, Frascati (Rome), Italy, EU}

\author{Xinran Wang}
\affiliation{Department of Astronomy, School of Physical Sciences, University of Science and Technology of China, Hefei 230026, China}

\author{Guanwen Fang}
\email{wen@mail.ustc.edu.cn}
\affiliation{School of Physics and Astronomy, Anqing Normal University, Anqing 246133, China}
\affiliation{Institute of Astronomy and Astrophysics, Anqing Normal University, Anqing 246133, China}

\author{Yi-Fu Cai}
\email{yifucai@ustc.edu.cn}
\affiliation{Department of Astronomy, School of Physical Sciences, University of Science and Technology of China, Hefei 230026, China}
\affiliation{CAS Key Laboratory for Research in Galaxies and Cosmology, School of Astronomy and Space Science, University of Science and Technology of China, Hefei 230026, China}


\begin{abstract}
Most dark matter direct-detection limits are reported under the
assumption of an isotropic, Maxwellian Standard Halo Model.
However, the Milky Way's stellar halo exhibits substantial radial
anisotropy, while cosmological simulations predict scale-dependent
anisotropy in the dark matter velocity distribution.
We investigate how departures from isotropy affect the
interpretation of LZ results in two complementary regimes:
background-dominated elastic-scattering searches and the recently
reported $248\,\keV$ nuclear-recoil candidate, which we study using
a background-free single-event approximation.
For elastic scattering, we use the self-consistent isotropic,
Osipkov--Merritt, axisymmetric, and triaxial halo models of
Belli et al.\ and translate the published limits through the ratio
of predicted event counts, applying a common escape-speed
truncation in the Galactic frame.
At WIMP masses above approximately $40\,\GeV$, the resulting
limits lie within about $\pm30\%$ of the isotropic reference.
Below approximately $15\,\GeV$, the model envelope broadens to
as much as an order of magnitude, reflecting sensitivity to the
high-velocity tail.
For the $248\,\keV$ event, the accessible inelastic mass splitting
is likewise controlled by this tail.
Some anisotropic models extend the allowed kinematic region
towards splittings at which solar-neutrino constraints on a
Higgsino interpretation become ineffective, although kinematic
accessibility alone does not establish viability at the fixed
Higgsino cross section.
These results show that halo-model dependence is modest at high
WIMP masses but becomes essential near kinematic thresholds.
\end{abstract}

\maketitle

\section{Introduction}
\label{sec:introduction}

Searches for weakly interacting massive particles (WIMPs) using
tonne-scale liquid-xenon detectors~\cite{AprileDoke2010} have
reached a sensitivity at
which astrophysical assumptions are increasingly important to
the interpretation of their
results~\cite{XENON100,LUX2017,PandaXII,XENON1T,PandaX4T,XENONnT,LZ22,Undagoitia2016,APPEC2022,Jungman1996,Roszkowski2018}.
The LZ experiment reported a $90\%$ confidence-level upper limit
on the spin-independent WIMP--nucleon cross section of
$9.2\times10^{-48}\,\cmSq$ at a WIMP mass of $36\,\GeV$, based on
an exposure of $0.9$ tonne-years~\cite{LZ22}.
A subsequent analysis using $4.2$ tonne-years obtained a limit of
$2.2\times10^{-48}\,\cmSq$ at $40\,\GeV$~\cite{LZ24}.
A separate search in an extended nuclear-recoil energy window,
using $2.84$ tonne-years, reported a candidate event with a
reconstructed recoil energy near $248\,\keV$~\cite{LZExtended2026}.

These results are commonly interpreted using a single assumed
Galactic dark matter phase-space distribution: the Standard Halo
Model (SHM), an isotropic Maxwellian distribution truncated at the
Galactic escape speed~\cite{Freese1988}.
The SHM provides a useful reference, but it specifies more than
a density profile: it also assumes an isotropic
velocity-dispersion tensor.

Stellar-halo observations motivate a closer examination of this
assumption~\cite{Evans:2018bqy}.
The stellar halo is radially biased, with representative
anisotropy parameters of approximately $0.4$--$0.7$ near the solar
radius~\cite{Belokurov18,Helmi2020}.
We use the convention
\[
\beta_{\mathrm{ani}}
=
1-\frac{\sigma_\theta^2+\sigma_\phi^2}{2\sigma_r^2}.
\]
This is a measurement of stellar tracers, rather than of dark
matter itself, and does not directly determine the dark matter
anisotropy.
Cosmological simulations nevertheless predict a scale-dependent
dark matter anisotropy that can modify the local high-velocity
tail~\cite{Wojtak13}.

At fixed local density and circular speed, velocity anisotropy
changes the relative populations of the bulk of the distribution
and its tail.
Direct-detection rates can therefore vary even when the local
dark matter density is held fixed.
For elastic spin-independent scattering with equal proton and
neutron couplings, the differential rate for a single isotope is
\begin{equation}
\begin{aligned}
&\frac{\dd R_A}{\dd \ER}
=
\frac{\rholoc\,\sigman}
     {2\mchi\mu_n^2}
A^2 F_A^2(\ER)\,\eta(\vmin),
\\
&\eta(\vmin,t)
=
\int_{|\bm v|>\vmin}
\frac{f_{\mathrm{lab}}(\bm v,t)}{|\bm v|}
\,\dd^3v .
\end{aligned}
\label{eq:basic_rate}
\end{equation}
Here $\mu_n$ is the WIMP--nucleon reduced mass, $F_A$ is the
nuclear form factor, and $f_{\mathrm{lab}}$ is normalized to unit
integral.
For elastic scattering,
$\vmin=\sqrt{m_A\ER/(2\mu_A^2)}$, where
$\mu_A=\mchi m_A/(\mchi+m_A)$.
Equation~\eqref{eq:basic_rate} assumes the standard
spin-independent contact interaction; the wider set of
nonrelativistic operators is reviewed in
Ref.~\cite{Fitzpatrick2013}.

At $\mchi\gtrsim40\,\GeV$, the recoil energies most relevant to LZ
typically probe $\vmin\sim200$--$300\,\kms$, where the halo models
considered here yield broadly similar predictions.
For lighter WIMPs, higher recoil energies, and especially
endothermic inelastic scattering, the required speed approaches
the largest available laboratory-frame speeds.
The predicted rate then becomes sensitive to the sparsely
populated high-velocity tail.

This distinction matters both when comparing experiments and
when interpreting candidate signals.
Experiments with different targets, thresholds, and energy
resolutions probe different ranges of $\vmin$, and proposed
ton-scale upgrades would extend this reach
further~\cite{LZprojection2020}.
Their relative sensitivities may therefore change when the SHM
assumption is relaxed.
The interpretation of anomalous observations, from the
DAMA/NaI annual modulation~\cite{BCFS,Bernabei:2013xsa,Savage2009,COSINE100}
to the high-energy LZ candidate~\cite{Graham,Freese}, is particularly
sensitive to such effects.

The high-velocity tail is also among the least constrained parts
of the local distribution.
Its prediction depends on the escape speed
~\cite{Smith07,Deason19,Roche24,Piffl2014,Monari2018}
and may be altered by non-equilibrium components, including
matter fluxes from the Large Magellanic Cloud
(LMC)~\cite{Graham,Erkal2019,Garavito2019}.

Two broad approaches are commonly used to assess the
astrophysical dependence of direct-detection constraints.
The first varies a small set of SHM parameters, such as the
local circular speed, escape speed, and dark matter
 density~\cite{McCabe2010}.
This captures parameter uncertainty within the SHM but retains
its functional form.
The second is the halo-independent approach, which replaces an
assumed distribution by a common halo integral
$\eta(\vmin)$~\cite{Fox}.
It is well suited to testing consistency between experiments
without specifying a halo model, but it does not directly assign
a limit to a particular astrophysical realization.

Self-consistent halo models offer a complementary approach.
They connect the velocity distribution to an underlying
gravitational model and allow its experimental consequences to
be evaluated explicitly.
In their study of the DAMA/NaI modulation, Belli, Cerulli,
Fornengo, and Scopel (BCFS) constructed four families of Galactic
models~\cite{Jaffe1983,NFW1996,Moore1999,Kravtsov1998,Evans2000,BCFS}:
spherical models with isotropic Eddington distribution functions;
the same density profiles with Osipkov--Merritt
anisotropy~\cite{Osipkov,Merritt};
axisymmetric Evans models~\cite{Evans93,Evans94};
and triaxial Gaussian models.
Their analysis showed that the inferred WIMP parameter region
depends substantially on the adopted halo.

Here, we use the BCFS suite as a controlled model set to assess
the halo-model dependence of LZ constraints.
We consider twenty-two realizations spanning four classes:
isotropic spherical models (class A), spherical
Osipkov--Merritt models (class B), axisymmetric models with and
without rotation (class C), and triaxial models (class D).
For each realization, we calculate the isotope-resolved recoil
spectrum, apply a Galactic-frame escape-speed truncation,
renormalize the velocity distribution, and fold the rate with
the adopted LZ efficiency.

For the background-dominated elastic searches, we translate the
published cross-section limit using
\begin{equation}
R(\mchi)
\equiv
\frac{\sigma_{\mathrm{model}}^{\mathrm{lim}}}
     {\sigma_{\SHM}^{\mathrm{lim}}}
\simeq
\frac{N_{\SHM}^{\mathrm{unit}}}
     {N_{\mathrm{model}}^{\mathrm{unit}}},
\label{eq:rescaling}
\end{equation}
where $N^{\mathrm{unit}}$ is the predicted event yield per unit
cross section. The $248\,\keV$ event requires a different treatment.
An elastic scatter on xenon deposits at most a few tens of keV
for the WIMP masses that dominate the LZ reach, so a single
$248\,\keV$ recoil is more naturally explained by endothermic
inelastic scattering, $\chi_1N\to\chi_2N$, in which a mass
splitting $\delta=m_{\chi_2}-m_{\chi_1}$ shifts the recoil
towards high energy~\cite{Freese,TuckerSmith2001}.
A thermal Higgsino~\cite{Krall:2017xij,Co:2021ion,m7qf-j8j6,Kowalska:2018toh}, with $m_\chi\simeq1.1\,\TeV$ and
splittings of a few hundred keV, is the most predictive
realization of this scenario~\cite{Graham:2024syw,Yin:2026jnn},
and its viability again hinges on the high-velocity tail.
For that analysis, we construct a single-event likelihood under
a background-free approximation rather than rescaling an
elastic exclusion curve.
The isotropic SHM and an LMC-like high-velocity tail serve as
reference distributions~\cite{Besla:2019xbx,Smith-Orlik:2023kyl,Donaldson_2022}.

Our results exhibit a clear hierarchy.
At high WIMP masses, halo modelling produces a modest correction
to the elastic limit.
At low masses, the same model variations are amplified by the
detector threshold.
For inelastic scattering, the tail can determine whether a
signal is kinematically accessible at all.

The remainder of the paper is organized as follows.
Section~\ref{sec:models} introduces the models and rate
calculation.
Sections~\ref{sec:si} and~\ref{sec:sd} present the
spin-independent and spin-dependent results.
Section~\ref{sec:inelastic} considers the high-energy event.
Section~\ref{sec:discussion} discusses the interpretation and
limitations, and Sec.~\ref{sec:conclusions} summarizes our
conclusions.

\section{Halo Models and Method}
\label{sec:models}

\subsection{Model suite}
\label{subsec:model_suite}

We adopt the four families of Galactic models constructed by
BCFS~\cite{BCFS}, summarized in Table~\ref{tab:models}.
Class A contains spherical models with isotropic distribution
functions.
Class B uses the same density profiles but introduces
Osipkov--Merritt anisotropy.
Class C consists of axisymmetric Evans models, which allow
flattening and rotation.
Class D comprises triaxial Gaussian distributions.

The isotropic model A0 is a Maxwellian with
$v_0=220\,\kms$ and $\vesc=544\,\kms$.
It defines the SHM reference used in
Eq.~\eqref{eq:rescaling}.

\begin{table}[t]
\caption{
Parameters of the halo models used in this study.
Classes A and B share the indicated density profiles, with
class B introducing Osipkov--Merritt anisotropy with
$\beta_0=0.4$.
The parameter $\beta$ in the density-profile entries is distinct
from the velocity-anisotropy parameter $\beta_{\mathrm{ani}}$.
For class D, $\delta_{\mathrm{DF}}$ denotes the distribution
parameter of Ref.~\cite{BCFS}, not the inelastic mass
splitting.
Speeds are in $\kms$ and length scales in $\kpc$.
}
\label{tab:models}
\scriptsize
\setlength{\tabcolsep}{3pt}
\begin{tabular}{
    @{}llp{0.47\columnwidth}@{}
}
\toprule
Model & Profile / DF & Parameters \\
\midrule
A0
& Maxwellian
& $v_0=220$, $\vesc=544$ \\

A1, B1
& Logarithmic
& $R_c=5$ \\

A2, B2
& Power law
& $R_c=16$, $\beta=0.7$ \\

A3
& Power law
& $R_c=2$, $\beta=-0.1$ \\

A4, B3
& Generalized
& $\alpha=1$, $\beta=4$, $\gamma=2$, $a=160$ \\

A5, B4
& Generalized
& $\alpha=1$, $\beta=3$, $\gamma=1$, $a=20$ \\

A6, B5
& Generalized
& $\alpha=1.5$, $\beta=3$, $\gamma=1.5$, $a=28$ \\

A7, B6
& Generalized
& $\alpha=2$, $\beta=3$, $\gamma=0.4$, $a=10$ \\

C1
& Evans, logarithmic
& $R_c=0$, $q=1/\sqrt{2}$ \\

C2
& Evans, logarithmic
& $R_c=5$, $q=1/\sqrt{2}$ \\

C3
& Evans, power law
& $R_c=16$, $q=0.95$, $\beta=0.9$ \\

C4
& Evans, power law
& $R_c=2$, $q=1/\sqrt{2}$, $\beta=-0.1$ \\

D1, D3
& Triaxial Gaussian
& $\delta_{\mathrm{DF}}=-1.78$,
  $p=1.013$, $q=1.024$ \\

D2, D4
& Triaxial Gaussian
& $\delta_{\mathrm{DF}}=16$,
  $p=1.013$, $q=1.024$ \\
\bottomrule
\end{tabular}
\end{table}

The spherical model suite includes logarithmic, power-law, and
generalized cusped profiles.
For the generalized profiles, we use
\begin{equation}
\rho(r)
=
\rholoc
\left(\frac{R_0}{r}\right)^\gamma
\left[
\frac{1+(R_0/a)^\alpha}
     {1+(r/a)^\alpha}
\right]^{(\beta-\gamma)/\alpha}.
\label{eq:density_profile}
\end{equation}
The logarithmic and power-law potential--density pairs are taken
from Ref.~\cite{BCFS}; they are treated separately rather than
identified with exact special cases of
Eq.~\eqref{eq:density_profile}.

For classes A and B, we obtain the gravitational potential by
direct integration of the spherical Poisson equation and
normalize the circular speed at the solar position to
$v_{\mathrm{rot}}(R_0)=220\,\kms$.
For the isotropic models, the distribution function follows
from Eddington inversion,
\begin{equation}
F(\mathcal{E})
=
\frac{1}{\sqrt{8}\,\pi^2}
\frac{\dd}{\dd\mathcal{E}}
\int_0^{\mathcal{E}}
\frac{\dd\Psi}{\sqrt{\mathcal{E}-\Psi}}
\frac{\dd\rho}{\dd\Psi},
\label{eq:eddington}
\end{equation}
where $\Psi$ is the relative potential and
$\mathcal{E}=\Psi-v^2/2$ is the relative energy.
The boundary prescription and numerical implementation follow
the adopted BCFS model construction.

For class B, the distribution function depends on energy and
angular momentum through
$Q=\mathcal{E}-L^2/(2r_a^2)$.
We use
$r_a=R_0\sqrt{(1-\beta_0)/\beta_0}$ with $\beta_0=0.4$.
The inversion has the same form as
Eq.~\eqref{eq:eddington}, with $\rho$ replaced by the augmented
density $(1+r^2/r_a^2)\rho$ and $\mathcal{E}$ replaced by
$Q$~\cite{Osipkov,Merritt}.

Class C uses the axisymmetric Evans distribution
functions~\cite{Evans93,Evans94}.
For the logarithmic halo, the velocity dependence includes a
factor proportional to
$(1+K v_\phi^2/v_0^2)\exp(-2v^2/v_0^2)$.
For the power-law halo, we use the three-term expression given
in Ref.~\cite{BCFS}.
The rotation prescription changes the relative populations of
prograde and retrograde orbits.

Class D uses triaxial Gaussian distributions,
\begin{equation}
f(\bm v)
\propto
\exp\left[
-\frac{v_r^2}{2\sigma_r^2}
-\frac{v_\theta^2}{2\sigma_\theta^2}
-\frac{v_\phi^2}{2\sigma_\phi^2}
\right],
\label{eq:triaxial}
\end{equation}
with dispersions determined by the adopted shape and
distribution parameters.
The major-axis and intermediate-axis branches correspond to
the two configurations specified by BCFS.
We retain their parameter choices to preserve a direct
comparison with the original DAMA analysis.

The suite is not a posterior distribution over Galactic
parameters.
It is a structured collection spanning isotropic, radially
biased, and tangentially biased velocity distributions, as well
as spherical, axisymmetric, rotating, and triaxial
configurations.
The resulting envelope should therefore be interpreted as a
model comparison, not a statistical confidence interval on the
Milky Way halo.

For each model, we record the fraction of the untruncated
distribution above the adopted escape speed.
This ranges from a few percent for the coldest models to
approximately $38\%$ for B4.
It measures how strongly a model is affected by the subsequent
cutoff; it is not an estimate of the true fraction of bound
Galactic particles above the escape speed.

The underlying distribution functions are self-consistent
before the common phenomenological truncation.
Imposing and renormalizing that cutoff does not, in general,
preserve exact global self-consistency with the original
potential.
We therefore use the truncated models as controlled local
velocity-distribution benchmarks.

\subsubsection{Numerical Construction of the Distribution Functions}
\label{app:numerics}

All models are constructed on a common radial grid of $500$
points spanning $10^{-2}$--$1000\,\kpc$.
For classes A and B, the gravitational potential is obtained by
direct integration and normalized at $R_0$.
Eddington or Osipkov--Merritt inversion is performed on the same
grid.

The derivative $\dd\rho/\dd\Psi$ is evaluated using cubic
interpolation.
The inversion integral is calculated by adaptive quadrature on
a grid of $200$ energies between zero and the maximum relative
potential.

The resulting distribution functions are sampled on a spherical
velocity grid with $300$ radial, $30$ polar, and $60$ azimuthal
points.
Elastic rates are averaged over twelve phases of the Earth's
orbit.

We validate the construction by reconstructing the density and
potential from the distribution function and monitoring
negative values introduced by the inversion.
The numerical screening used for this model set rejects
realizations with a negative fraction above $10\%$, leaving
twenty-two retained models.

This threshold is a numerical screening rule, not a physical
allowance for a negative distribution function.
A physical distribution must be non-negative.
Residual negative values must therefore be checked for
convergence and shown to be numerical artifacts before the
corresponding realization is regarded as validated.

For every model, we also record the fraction above the adopted
escape speed before truncation.
This ranges from a few percent for the coldest models to
approximately $38\%$ for B4.
It characterizes sensitivity to the cutoff rather than the
measured population of unbound dark matter.

The escape-speed truncation is subsequently applied in the
Galactic frame and the local distribution is renormalized.
All quoted halo ratios use the same local dark matter density,
so that the comparison isolates changes in the velocity
weighting rather than an independent density-normalization
uncertainty.

\subsection{Truncation, rate calculation, and limit rescaling}
\label{subsec:rate_method}

We impose the cutoff in the Galactic frame.
Writing $\bm v$ for the laboratory-frame velocity,
\begin{equation}
\begin{aligned}
f_{\mathrm{lab}}^{\mathrm{tr}}(\bm v,t)
={}&
\frac{
f_{\mathrm{gal}}\!\left(\bm v+\bm v_{\mathrm{lab}}(t)\right)
}{
\mathcal{N}(t)
}
\\
&\times
\Theta\!\left[
\vesc-\left|\bm v+\bm v_{\mathrm{lab}}(t)\right|
\right],
\end{aligned}
\label{eq:truncation}
\end{equation}
where $\mathcal{N}$ normalizes the distribution to unit integral
and
$\bm v_{\mathrm{lab}}(t)=\bm v_\odot+\bm v_\oplus(t)$.
We adopt
$\bm v_\odot=(11.1,232.24,7.25)\,\kms$.
Elastic event yields use the annual average over the sampled
Earth-orbit phases.

The baseline is $\vesc=544\,\kms$.
For classes A and B, we vary it between $504$ and
$584\,\kms$, covering the local estimate
$528^{+24}_{-25}\,\kms$~\cite{Deason19}.
More recent determinations further motivate treating the escape
speed as a systematic rather than an exact input
~\cite{Roche24}.

For classes C and D, we additionally extend the imposed
truncation to $1000\,\kms$.
This is a sensitivity test of the model tail, not an alternative
measurement of the Galactic escape speed.
A physical interpretation of such an extension would require
an additional high-speed component or a modified dynamical
description.

For a multi-isotope xenon target, the spin-independent rate is
\begin{equation}
\frac{\dd R}{\dd\ER}
=
\frac{\rholoc\,\sigman}{2\mchi\mu_n^2}
\sum_A
w_A\,\mathcal{C}_A^2
F_A^2(\ER)\,
\eta\!\left[\vmin(\ER,m_A)\right],
\label{eq:isotope_rate}
\end{equation}
where $w_A$ is the isotope mass fraction.

For the standard isospin-conserving SI limits in
Sec.~\ref{sec:si}, $\mathcal{C}_A=A$.
For the $Z$-mediated higgsino calculation in
Sec.~\ref{sec:inelastic}, we instead use the weak-charge factor
$\mathcal{C}_A=Q_A$ with an explicitly specified reference
cross-section convention.
The same interaction convention must be used in the numerator
and denominator of Eq.~\eqref{eq:rescaling}.
These two coupling choices should not be mixed when
interpreting a published WIMP--nucleon limit.

We use the Helm form factor~\cite{Helm1956,LewinSmith1996,Duda2007}
and sum over the nine xenon isotopes.
Integrating the recoil spectrum over the detector efficiency
gives the expected yield,
\[
N
=
\mathcal{E}_{\mathrm{exp}}
\int \dd\ER\,
\eff(\ER)\,
\frac{\dd R}{\dd\ER},
\]
where $\mathcal{E}_{\mathrm{exp}}$ is the exposure.

The event-yield rescaling assumes that the data approximately
constrain the same maximum number of additional signal events
for each halo model.
The cross-section limit then varies inversely with the predicted
yield per unit cross section.

Setting the model to A0 gives $R=1$ by construction and provides
an implementation check.
It does not establish the accuracy of the rescaling
approximation.
Changes in the recoil spectrum can affect signal--background
discrimination and hence the fitted signal upper bound.
We therefore interpret the resulting envelopes as halo-model
systematic variations on the published limits, rather than
exact likelihood-derived exclusions.

\section{Spin-Independent Limits}
\label{sec:si}

\begin{figure*}[t]
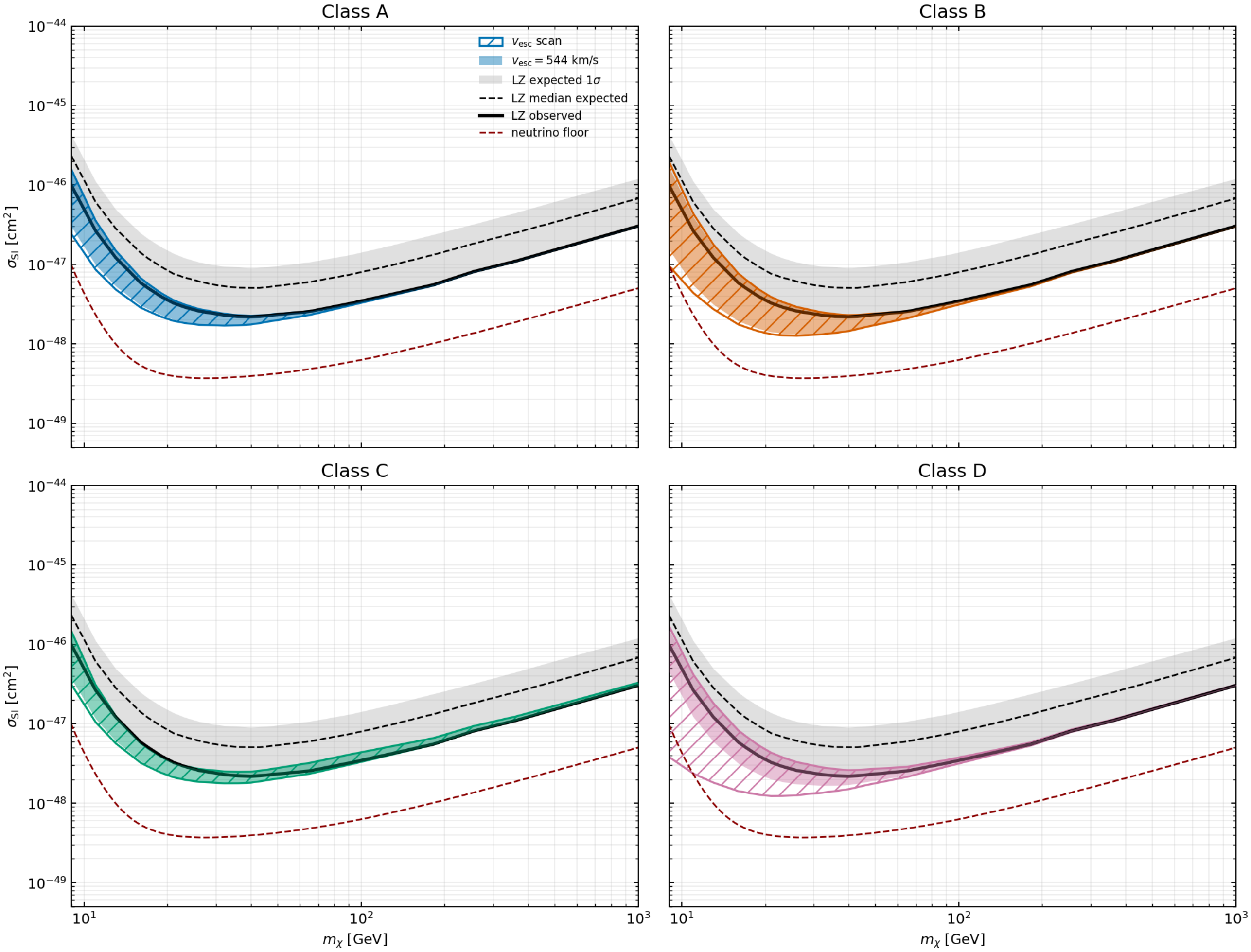

\centering
\SafeIncludeGraphics[width=0.96\textwidth]
{\SIFigurePath}
{Spin-independent limits for the four halo families.}
\caption{
Spin-independent limits for the four halo families, obtained by
rescaling the published LZ limit using
Eq.~\eqref{eq:rescaling}.
The black curve shows the observed LZ limit.
The light-gray band denotes the expected
$\pm1\sigma$ range around the median expected limit.
The baseline model envelope is evaluated at
$\vesc=544\,\kms$, while the coloured hatched region shows
the escape-speed scan.
The red dash-dotted curve is the neutrino-floor benchmark.
The halo-model envelopes represent systematic model variations,
not statistical confidence bands.
}
\label{fig:si}
\end{figure*}

Figure~\ref{fig:si} shows the rescaled SI limits for the four halo
families.
The results divide naturally into a high-mass regime, where the
model dependence is modest, and a low-mass regime, where it
becomes substantial.

At $\mchi\gtrsim40\,\GeV$, the models span approximately
$R=0.8$--$1.3$.
The spread is less than a factor of two and is comparable to
the experimental uncertainty.
Towards lower masses, the envelope broadens as the minimum speed
required for an above-threshold recoil approaches the largest
available speeds.
At $9\,\GeV$, the ranges are
$R=0.32$--$1.48$ for class A,
$0.17$--$1.63$ for class B,
$0.39$--$1.37$ for class C, and
$0.45$--$1.53$ for class D.
Near $1\,\TeV$, the combined range narrows to approximately
$R=0.89$--$1.37$.

\begin{figure}[t]
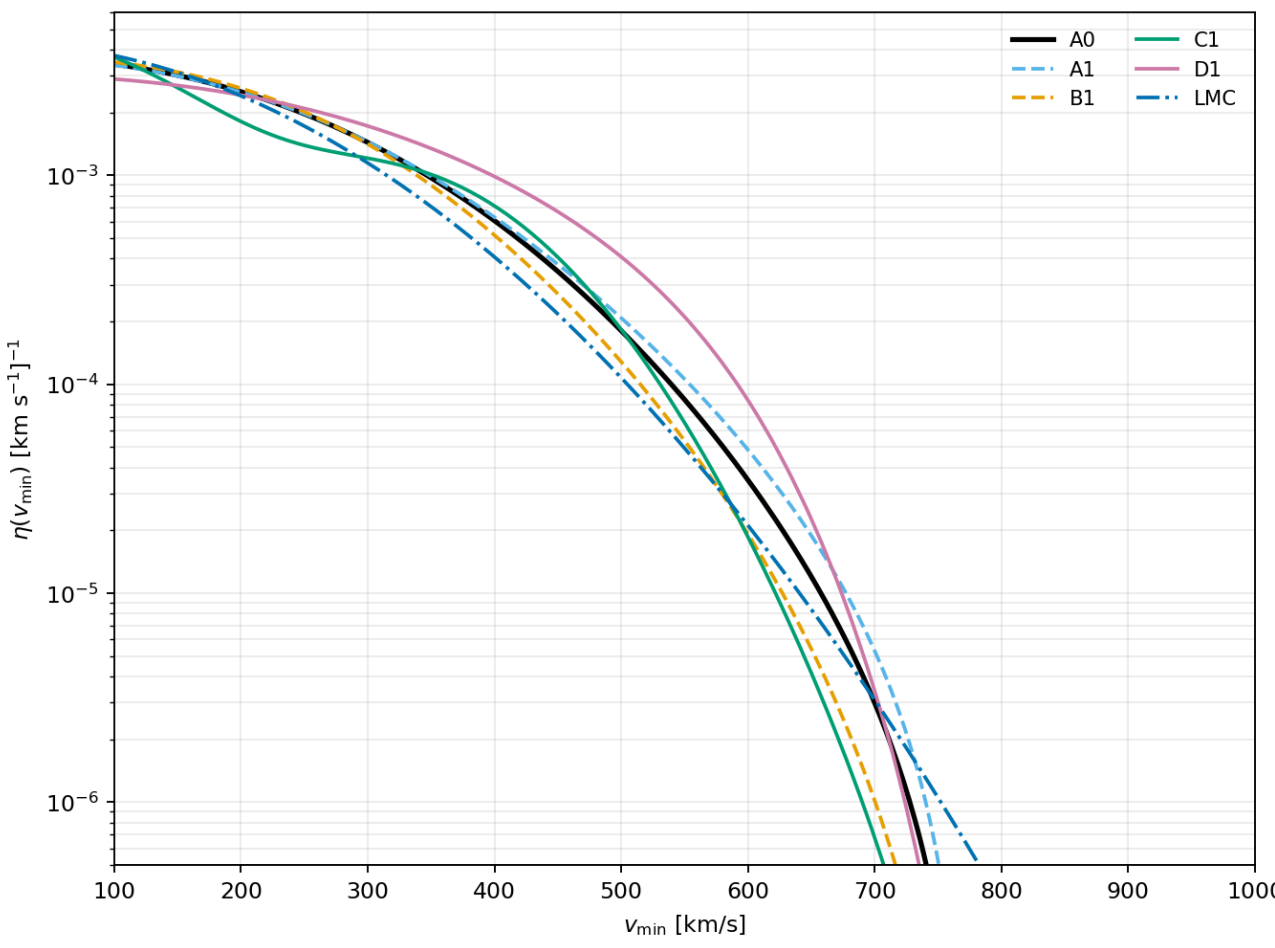

\centering
\SafeIncludeGraphics[width=\linewidth]
{\HaloIntegralFigurePath}
{Halo integral for representative models}
\caption{
Halo integral $\eta(\vmin)$ for representative models:
the isotropic baseline A0, the cold models B1 and C1,
a radially biased model D1, and an LMC-like high-velocity tail.
The equilibrium reference models use a Galactic-frame cutoff
of $\vesc=544\,\kms$.
The predictions are similar at
$\vmin\simeq200\,\kms$, where the bulk of the distribution
dominates, but differ by factors of two to five at
$\vmin\gtrsim500\,\kms$.
Neglecting the Earth's orbital contribution, the corresponding
maximum laboratory-frame speed for the baseline cutoff is
$\vesc+|\bm v_\odot|\simeq776\,\kms$.
The LMC-like reference extends to higher speeds.
}
\label{fig:halo_integral}
\end{figure}

Figure~\ref{fig:halo_integral} shows the origin of the mass
dependence.
At $\vmin\sim200$--$300\,\kms$, the range most relevant near
$40\,\GeV$, the halo integrals differ by only tens of percent.
At $\vmin\gtrsim500\,\kms$, which is increasingly important for
light WIMPs, they differ by factors of two to five.
The widening low-mass envelope therefore follows from
differences in the number of particles capable of producing
detectable recoils.

The ordering of the limits is systematic.
Models with more populated high-velocity tails produce larger
event yields and hence smaller values of $R$.
These include the generalized profiles A5--A7 and B4--B6, as well
as radially biased models in classes C and D.
Conversely, the cold models B1 and C1 yield weaker constraints
because fewer particles occupy the required speed range.

The pre-truncation fraction above $544\,\kms$ is a useful
diagnostic but does not determine the limit on its own.
The rate depends on the retained, renormalized distribution over
the full range of $\vmin$ weighted by the detector response.
A large removed fraction can therefore correlate with, but is
not equivalent to, a large retained signal rate.

The escape-speed scan introduces an additional uncertainty that
is small at high masses and increases near threshold.
At $\mchi\gtrsim40\,\GeV$, changing $\vesc$ from
$504$ to $584\,\kms$ modifies the limits by approximately
$5\%$ or less.
Extending the cutoff to $1000\,\kms$ for classes C and D changes
them by only $10$--$20\%$.

At $9\,\GeV$, these variations become much more important.
For classes A and B, the escape-speed scan broadens the band by
approximately a factor of two.
For classes C and D, the effect can be larger because the
extended cutoff is a prescribed tail parameter rather than a
measured uncertainty within a fixed equilibrium model.
Within class-D realization, D3, extending the cutoff
from $544$ to $1000\,\kms$ strengthens the limit by approximately
a factor of ten, lowering the envelope to $R\simeq0.04$.

This sensitivity cannot be represented by uncertainty in the
density profile alone.
It reflects the combined influence of the velocity
distribution, the cutoff, and the detector threshold.

We also vary the efficiency treatment.
The tabulated response adopted for the elastic calculation
covers $0$--$100\,\keV$, rises from zero near
$2\,\keV$, reaches a broad plateau between approximately
$5$ and $40\,\keV$, and returns to zero at $100\,\keV$.
We interpolate it linearly and set it to zero outside the
tabulated interval.

As a sensitivity test, we replace this response with a step
function of unit efficiency above threshold.
This changes the high-mass limits by approximately $10$--$20\%$,
but by as much as a factor of two to three at $9\,\GeV$, where
the recoil spectrum is concentrated near threshold.
The low-mass interpretation is thus controlled jointly by the
high-velocity tail and the detector response.

Table~\ref{tab:si_envelopes} summarizes the baseline envelopes.

\begin{table}[t]
\caption{
Envelope of the rescaled SI limit,
$R=\sigma_{\mathrm{model}}^{\mathrm{lim}}/
\sigma_{\SHM}^{\mathrm{lim}}$,
for each halo family at selected WIMP masses.
The baseline cutoff is $\vesc=544\,\kms$.
}
\label{tab:si_envelopes}
\centering
\small
\setlength{\tabcolsep}{4pt}
\begin{tabular}{@{}ccccc@{}}
\toprule
$\mchi\,[\GeV]$ & A & B & C & D \\
\midrule
9
& $0.32$--$1.48$
& $0.17$--$1.63$
& $0.39$--$1.37$
& $0.45$--$1.53$ \\

40
& $0.91$--$1.01$
& $0.84$--$1.01$
& $0.92$--$1.27$
& $0.87$--$1.14$ \\

1000
& $0.98$--$1.15$
& $0.98$--$1.37$
& $0.89$--$1.12$
& $0.91$--$1.19$ \\
\bottomrule
\end{tabular}
\end{table}

\section{Spin-Dependent Limits}
\label{sec:sd}

We apply the same event-yield procedure to spin-dependent
scattering.
The rate remains proportional to the halo integral
$\eta(\vmin)$, while the coherent SI nuclear response is replaced
by the appropriate spin structure function.
This changes the recoil-energy weighting, but not the
rescaling argument for background-dominated data.

\begin{figure*}[t]
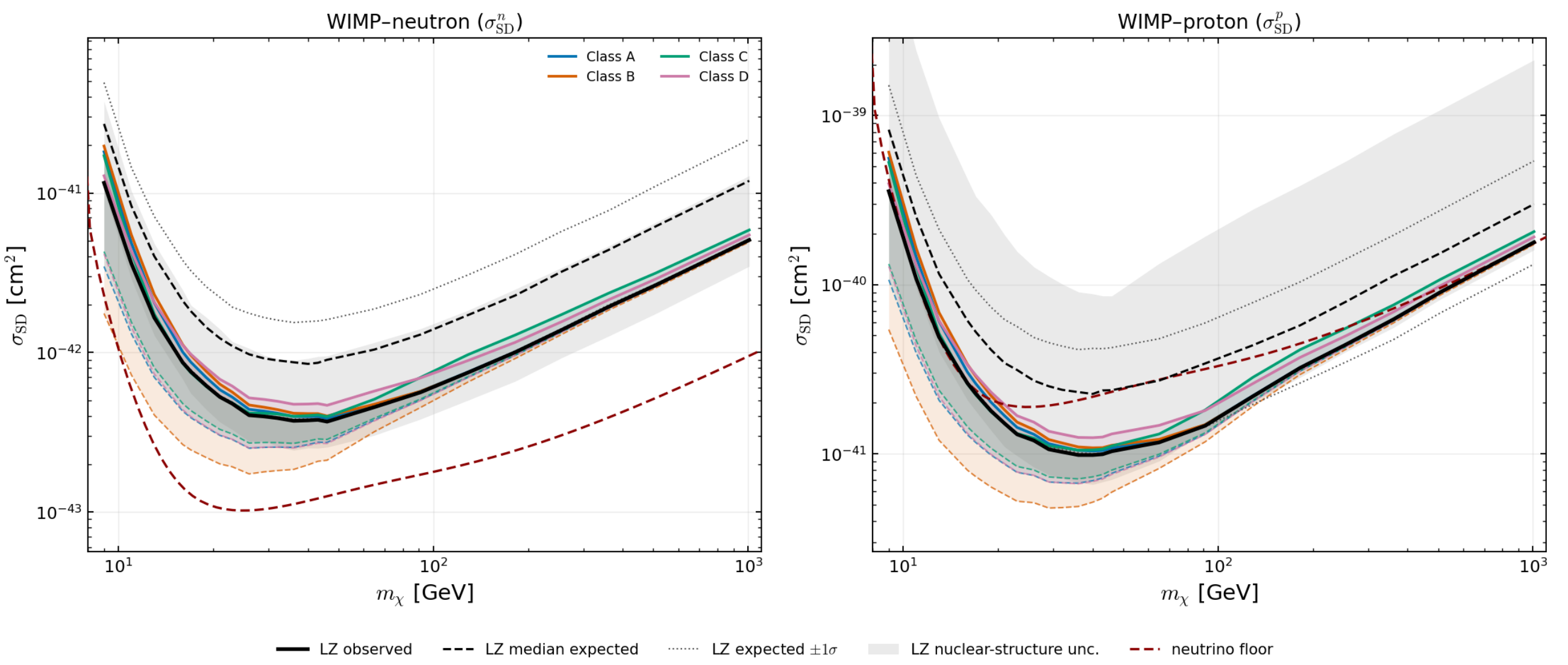

\centering
\SafeIncludeGraphics[width=0.96\textwidth]
{\SDFigurePath}
{Spin-dependent WIMP--neutron and WIMP--proton limits.}
\caption{
Spin-dependent limits for WIMP--neutron scattering (left)
and WIMP--proton scattering (right).
The black solid and dashed curves show the observed and median
expected LZ limits, respectively.
Dotted curves indicate the $\pm1\sigma$ expected range.
The gray band represents nuclear-structure uncertainty.
The red dashed curve shows the neutrino-floor
benchmark~\cite{OHare21}.
Coloured curves delimit the halo-model envelopes for the four
families.
}
\label{fig:sd}
\end{figure*}

Figure~\ref{fig:sd} shows the resulting WIMP--neutron and
WIMP--proton limits.
Their halo-model dependence resembles that of the SI results.
At $\mchi\simeq40\,\GeV$, the model ranges are
$R=0.79$--$1.03$ for class A,
$0.65$--$1.06$ for class B,
$0.82$--$1.13$ for class C, and
$0.78$--$1.26$ for class D.
At $9\,\GeV$, these broaden to
$0.31$--$1.50$,
$0.16$--$1.65$,
$0.38$--$1.40$, and
$0.37$--$1.13$, respectively.

The strongest constraints again arise from models with more
populated high-velocity tails, including B4, B6, D1, and D3,
whereas B1 and C1 give the weakest limits.
This similarity is expected because both SI and SD rates probe
the same halo integral.

The residual differences arise from the nuclear response.
The adopted SD structure functions change the weighting of
large momentum transfers and hence the sensitivity to the
highest-$\vmin$ recoils~\cite{Menendez2012}.
For example, at $9\,\GeV$ the class-D SD envelope,
$0.37$--$1.13$, is narrower than the corresponding SI envelope,
$0.45$--$1.53$.

Two features of the published SD results are important for their
interpretation.
First, the observed limits lie below the median expected limits
in both channels.
At $40\,\GeV$, the observed WIMP--neutron limit is
$3.75\times10^{-43}\,\cmSq$, compared with a median expectation
of $8.48\times10^{-43}\,\cmSq$ and a lower one-standard-deviation
value of $3.84\times10^{-43}\,\cmSq$.
For WIMP--proton scattering, the corresponding values are
$9.82\times10^{-42}$,
$2.26\times10^{-41}$, and
$1.00\times10^{-41}\,\cmSq$.
Both observed limits are therefore close to the lower edge of
the expected $1\sigma$ band.

Experimental reach should consequently be assessed using the
expected sensitivity as well as the observed limit.
An unusually strong observed exclusion does not imply an
equally strong median sensitivity.

Second, the two channels occupy different positions relative
to the neutrino-floor benchmark~\cite{OHare21,Billard2014,Ng2017}.
At $40\,\GeV$, the observed neutron limit lies approximately a
factor of $3.2$ above the quoted benchmark of
$1.19\times10^{-43}\,\cmSq$, while the median expectation lies
approximately a factor of seven above it.
For proton scattering, the median expectation is close to the
benchmark,
$2.26\times10^{-41}$ versus
$2.20\times10^{-41}\,\cmSq$,
whereas the observed limit lies a factor of $2.2$ below it.

There is no inconsistency in an exclusion limit falling below
such a benchmark.
The neutrino floor characterizes discovery sensitivity under
specified assumptions; it is not a hard lower boundary on
exclusion limits.

Near $40\,\GeV$, halo modelling is not the dominant theoretical
uncertainty in the SD channels.
The nuclear-structure band spans
$2.5\times10^{-43}$--$9.2\times10^{-43}\,\cmSq$
for neutron scattering, a factor of $3.6$, and
$6.6\times10^{-42}$--$8.8\times10^{-41}\,\cmSq$
for proton scattering, a factor of approximately $13$.
By comparison, the halo-model envelope spans approximately a
factor of two.
Nuclear-response uncertainty is therefore larger in this mass
range, although halo uncertainty grows substantially towards
low WIMP masses.

\section{Inelastic Scattering and the 248 keV Event}
\label{sec:inelastic}

The LZ extended-energy search reported a candidate with
reconstructed nuclear-recoil energy
$\ER=248\pm23_{\mathrm{stat}}\pm23_{\mathrm{sys}}\,\keV$
in an exposure of $2.84$ tonne-years~\cite{LZExtended2026}.
The event occurred in June, near the seasonal maximum of the
Earth's speed relative to the Galactic frame.
It lies near the upper part of the extended recoil-energy
acceptance, which reaches approximately $270\,\keV$.

The event is noteworthy because of its high recoil energy and
small expected background, but it does not establish a dark
matter discovery yet.
The LZ analysis reports a maximum local significance of about
$3.4\sigma$ across the models tested, and a global significance of
about $2.6\sigma$ once the look-elsewhere effect is
included~\cite{LZExtended2026}.
Any interpretation must account for the detector response,
background modelling, and the absence of the additional
lower-energy events predicted by many elastic models.

For conventional coherent elastic scattering, the recoil
spectrum falls rapidly with energy.
Explaining an isolated $248\,\keV$ recoil with the same
interaction responsible for the standard SI signal would
generally overpredict lower-energy recoils.
This statement concerns the conventional elastic interaction
considered here, rather than all possible elastic operators.

Endothermic inelastic scattering offers a different kinematic
possibility~\cite{TuckerSmith2001,Chang2009,Gould1987,Nussinov2009,Blennow2016,Catena2018}.
If the incoming state $\chi_1$ scatters into a heavier state
$\chi_2$, a fraction of its kinetic energy is used to overcome
the mass splitting.
The resulting threshold can suppress lower-energy events and
shift the observable recoil spectrum towards higher energies.

Our purpose is not to claim that the LZ event is a dark matter
signal.
We ask which combinations of halo model, mass splitting, and
cross section are compatible with it under a specified
single-event likelihood, and how that compatibility depends on
the high-velocity tail.

For
$\chi_1 N\rightarrow\chi_2 N$
with $\delta=m_{\chi_2}-m_{\chi_1}>0$, the minimum speed required
to produce a recoil $\ER$ is
\begin{equation}
\vmin(\ER)
=
\frac{1}{\sqrt{2m_A\ER}}
\left(
\frac{m_A\ER}{\mu_A}+\delta
\right).
\label{eq:vmin_inelastic}
\end{equation}
Its minimum with respect to recoil energy is
\begin{equation}
\vmin^{\mathrm{min}}
=
\sqrt{\frac{2\delta}{\mu_A}},
\qquad
\ER^\star=\frac{\mu_A\delta}{m_A}.
\label{eq:vmin_minimum}
\end{equation}

For a largest available laboratory-frame speed $\vmax$, this
implies the necessary kinematic condition
$\delta\leq\mu_A\vmax^2/2$.
The high-velocity tail therefore controls not only the rate but
also the maximum accessible splitting.
The frame in which a quoted tail endpoint is defined must be
specified before applying this condition.

We calculate the isotope-resolved rate using
\begin{equation}
\begin{aligned}
\frac{\dd R}{\dd\ER}
&=
\frac{\rholoc\,\sigman}
     {2\mchi\mu_n^2}
\sum_A w_A Q_A^2 F_A^2(\ER)
\eta\!\left[\vmin(\ER)\right],
\\
Q_A
&=
N_A-\left(1-4\sin^2\theta_W\right)Z_A ,
\end{aligned}
\label{eq:inelastic_rate}
\end{equation}
where $N_A$ and $Z_A$ are the neutron and proton numbers.
In this equation, $\sigman$ denotes the reference cross section
in the weak-charge normalization used for the higgsino
benchmark.
It should not be identified with an isospin-conserving SI
cross section without specifying the coupling conversion.

The expected event count and the differential event weight at
the observed energy are
\begin{equation}
\begin{aligned}
N(\delta,\sigman)
&=
\mathcal{E}_{\mathrm{exp}}
\int \dd\ER\,
\frac{\dd R}{\dd\ER}\,
\eff(\ER),
\\
B(\Eobs;\delta,\sigman)
&=
\mathcal{E}_{\mathrm{exp}}
\int \dd\ER\,
\frac{\dd R}{\dd\ER}\,
\eff(\ER)\,
G(\Eobs\mid\ER).
\end{aligned}
\label{eq:count_and_weight}
\end{equation}
Here $\mathcal{E}_{\mathrm{exp}}=2.84$ tonne-years.
The function $G$ is a normalized Gaussian response with
width
\begin{equation}
\sigma_E(\ER)
=
\sqrt{
(23\,\keV)^2
\frac{\ER}{248\,\keV}
+
\left(0.093\,\ER\right)^2
}.
\label{eq:energy_resolution}
\end{equation}

Equation~\eqref{eq:energy_resolution} is an approximate
smearing prescription that combines the quoted statistical and
systematic energy scales in quadrature.
It is not a substitute for a detector likelihood with an
explicit energy-scale nuisance parameter.

For consistency with the benchmark calculation, we adopt unit
efficiency over the high-energy range that contributes to the
signal and verify that increasing the upper integration
boundary does not change the result once the relevant recoil
support is included.
This is an explicit approximation.
The extended-search response is not determined by the
$0$--$100\,\keV$ efficiency table used for the elastic calculation.
A definitive reinterpretation should use the published
extended-energy acceptance and selection model.

The statistical treatment differs from the elastic case.
We use a background-free approximation and an extended
single-event likelihood,
\begin{equation}
\begin{aligned}
\ln L(\sigman,\delta)
={}&
-\sigman N_{\mathrm{unit}}(\delta)
\\
&+
\ln\!\left[
\sigman
B_{\mathrm{unit}}(\Eobs;\delta)
\right]
+\mathrm{const.},
\end{aligned}
\label{eq:single_event_likelihood}
\end{equation}
where $N_{\mathrm{unit}}$ and $B_{\mathrm{unit}}$ are evaluated
per unit cross section.
The symbol $B$ denotes the signal event-density weight, not a
background contribution.

We define nominal $90\%$ likelihood contours by
$-2\Delta\ln L=4.61$, or equivalently
$\Delta\ln L\simeq-2.30$.
This is the usual two-parameter asymptotic
threshold~\cite{Wilks1938,Cowan2011}; how such intervals behave
when the signal is small is discussed in
Ref.~\cite{FeldmanCousins1998}.
Because only one event is used and the parameter space contains
kinematic boundaries, its exact frequentist coverage is not
guaranteed.
We therefore refer to these regions as nominal likelihood
contours rather than coverage-calibrated confidence regions.

\begin{figure*}[t]
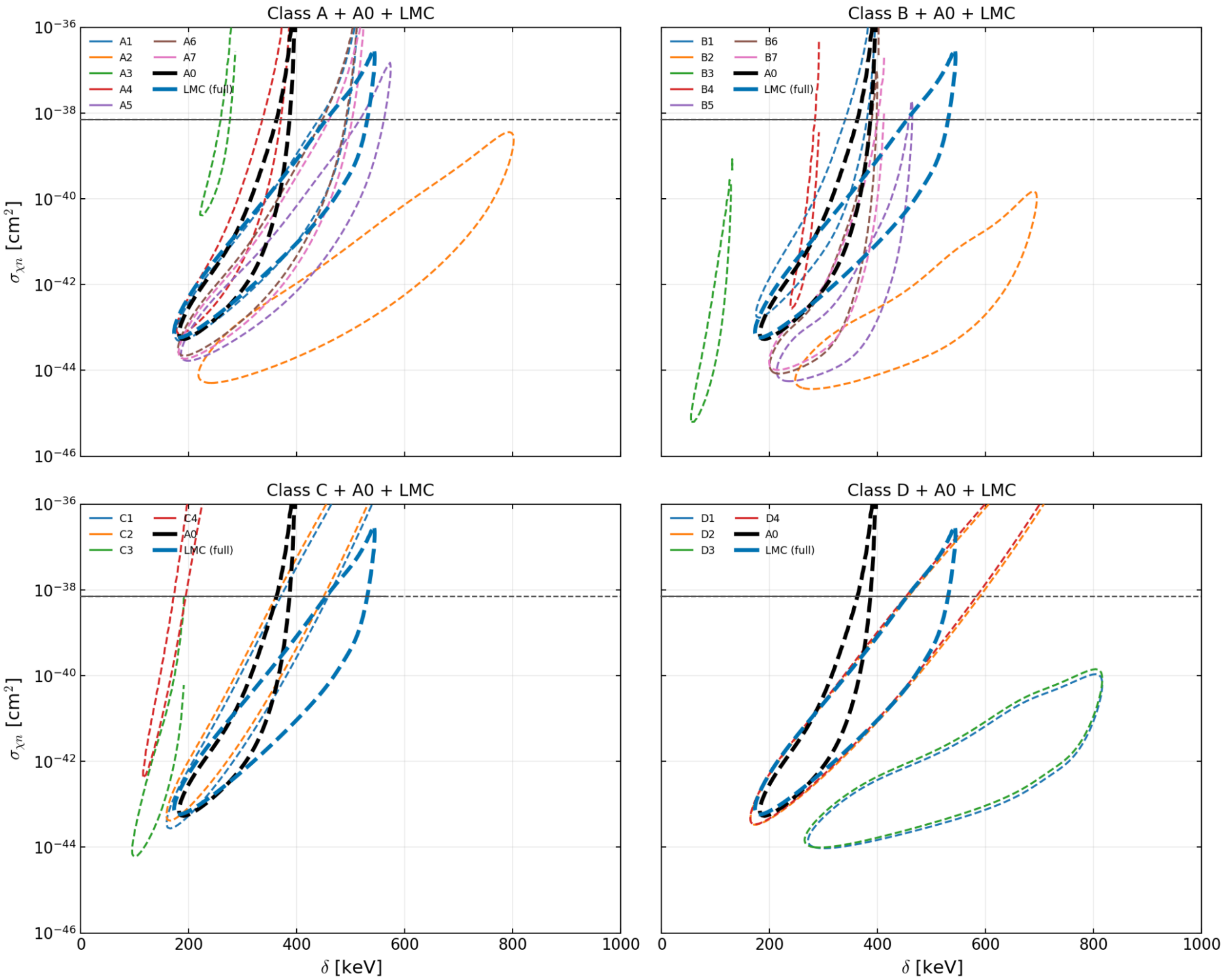

\centering
\SafeIncludeGraphics[width=0.96\textwidth]
{\InelasticFigurePath}
{Single-event likelihood contours for inelastic scattering.}
\caption{
Nominal $90\%$ likelihood contours in the
$(\delta,\sigman)$ plane for $\mchi=1.1\,\TeV$.
The isotropic reference A0 and the LMC-like tail are shown as
thick black and blue curves, respectively.
Thin curves represent the individual models in each family.
The horizontal line denotes the higgsino benchmark cross section
in the adopted weak-charge normalization.
It is drawn solid where solar-neutrino observations exclude the
higgsino interpretation ($\delta\lesssim566\,\keV$) and dashed
where that constraint no longer applies.
The change of style marks the splitting beyond which inelastic
capture in the Sun is kinematically suppressed, so that the
solar-neutrino limit weakens and eventually disappears.
The Galactic-frame cutoff is $544\,\kms$ for classes A and B
and $1000\,\kms$ for classes C and D.
The latter is a tail-sensitivity test rather than a measured
escape-speed uncertainty.
}
\label{fig:inelastic}
\end{figure*}

Figure~\ref{fig:inelastic} shows the contours for the isotropic
reference, the LMC-like tail, and the four halo families.
The dominant feature is the kinematic ceiling.
In the adopted numerical realizations, the reported maximum
splitting increases from approximately $384\,\keV$ for the SHM
reference to approximately $556\,\keV$ for the LMC-like tail.

At $\delta=350\,\keV$, producing a $248\,\keV$ xenon recoil
requires a speed of approximately $760\,\kms$.
Only a small fraction of the SHM distribution occupies this
laboratory-frame range.
The contours are consequently sensitive to modest changes in
the tail.

The anisotropic models bracket the reference distributions.
Classes A, B, and C typically close between approximately
$400$ and $550\,\keV$, while class D can extend further.
For the radially biased models D1 and D3, the allowed contours
span approximately $240$--$700\,\keV$, with their high-splitting
ends occurring near
$\sigman\simeq3\times10^{-40}\,\cmSq$.
The tangentially biased models D2 and D4 span approximately
$165$--$775\,\keV$, but require larger cross sections because
their event rates are more strongly suppressed.

As in the elastic case, the ordering is controlled by the halo
integral at large $\vmin$.
Its effect is amplified by the inelastic threshold:
a factor-of-two change in $\eta$ near
$\vmin\sim700\,\kms$ can shift the accessible splitting by
several tens of keV.

A nearly pure thermal higgsino provides a useful particle-physics
benchmark~\cite{Freese,NagataShirai2015}.
In the heavy-gaugino limit, the relic abundance selects a mass
near $1.1\,\TeV$, and the off-diagonal $Z$ coupling fixes the
inelastic scattering strength.
The neutral-state splitting, however, depends on the
supersymmetric mass parameters and their radiative corrections;
it is not fixed to a few hundred keV by the relic abundance
alone.
The $340$--$360\,\keV$ range should therefore be understood as a
benchmark region motivated by fits to the event, rather than a
universal prediction of a thermal higgsino.

In the normalization adopted in Ref.~\cite{Graham}, the
higgsino scattering cross section is approximately
$\sigman\simeq7\times10^{-39}\,\cmSq$.
This fixed interaction strength makes the interpretation
predictive but also strongly constrained.

For the benchmark assumptions used in the cited analyses,
solar-neutrino observations exclude the relevant higgsino
parameter space for splittings below approximately
$566\,\keV$~\cite{Pospelov:2026ewn,IceCube2017,Cohen2013}.
The SHM-preferred range near $300$--$400\,\keV$ therefore lies
within the excluded region.

Increasing the high-speed population raises the rate at fixed
$(\delta,\sigman)$ and can move the preferred splitting upwards.
In Fig.~\ref{fig:inelastic}, the higgsino cross section intersects
the SHM contour near $\delta\simeq360\,\keV$ and the LMC-like
contour near $\delta\simeq455\,\keV$.
Both remain below the quoted solar-neutrino threshold.
The more extended tails discussed in
Ref.~\cite{Graham} can shift the preferred region further,
but require an additional population at very high speeds.

The distinction between kinematic accessibility and
particle-model viability is essential.
A contour may extend beyond $566\,\keV$ only at cross sections
well below the fixed higgsino value.
Such an extension does not by itself evade the higgsino
constraint.

The inelastic interpretation also has testable consequences.
Radiative decay,
$\chi_2\rightarrow\chi_1\gamma$,
would produce a photon with energy close to the splitting.
For a sufficiently long decay length, a large-volume detector
could search for this photon independently of the original
nuclear recoil.

Measurements using different target nuclei would probe the
target-mass dependence of the inelastic threshold and help break
the degeneracy between $\mchi$ and $\delta$.
Heavy-element paleodetectors would additionally probe a
time-integrated dark matter distribution, potentially testing
whether the present-day tail is associated with the
LMC~\cite{Graham:2026ivn}.
A larger xenon exposure could test the recoil spectrum and
seasonal dependence through additional events or constrain the
interpretation if no corresponding signal appears.

The high-energy event thus illustrates a reversal of the usual
role of halo modelling.
For high-mass elastic scattering, the halo mainly changes the
normalization of an existing constraint.
For inelastic scattering, it can control whether the proposed
signal is possible at all.

\section{Discussion}
\label{sec:discussion}

The results identify three regimes of halo-model dependence.

First, high-mass elastic-scattering limits are comparatively
robust.
Above approximately $40\,\GeV$, the four halo families yield
constraints within about $\pm30\%$ of the isotropic reference.
The relevant recoils probe the bulk of the velocity
distribution, where the halo integrals are similar.

Second, low-mass limits are substantially more model-dependent.
Below approximately $15\,\GeV$, detectable recoils require
particles in the high-velocity tail, and the envelope broadens
to as much as an order of magnitude.
The relevant quantity is the population remaining above the
detector's kinematic threshold after truncation and
renormalization, rather than the density profile alone.

Third, endothermic scattering depends still more directly on
the tail.
The maximum accessible splitting is fixed by the largest
available speeds.
For the $1.1\,\TeV$ benchmark, the adopted reference
distributions give ceilings of approximately
$384$ and $556\,\keV$.
Some anisotropic models extend the kinematic reach further,
but the extension need to not occur at the fixed Higgsino cross
section.

The relative importance of uncertainties also depends on the
interaction.
Near $40\,\GeV$, the SD halo-model envelope spans approximately
a factor of two, whereas the nuclear-response uncertainty spans
factors of about $3.6$ and $13$ in the neutron and proton
channels.
Improved nuclear calculations are therefore needed before halo
modelling becomes the leading theoretical uncertainty in that
mass range.

Several qualifications limit the interpretation of the
envelopes.
The elastic results use event-yield rescaling, not a complete
fit to the LZ likelihood.
They estimate halo dependence but do not capture all effects of
a changed signal spectrum on background discrimination.

The BCFS suite is a structured model set rather than a
statistically weighted Galactic posterior.
Moreover, its members differ in density profile, geometry,
rotation, and velocity anisotropy.
The full envelope therefore measures combined halo-model
dependence, not anisotropy alone.
Matched comparisons between class-A and class-B models provide
a cleaner way to isolate the effect of the adopted
Osipkov--Merritt anisotropy.

The high-speed extensions of classes C and D are sensitivity
tests, not confidence intervals on the measured escape speed.
A physical interpretation would require an explicit source and
normalization for the additional tail population.
The common cutoff also changes the globally self-consistent
distribution from which each local model was derived.

Finally, the single-event contours depend on an approximate
energy response, a unit-efficiency prescription, and a
background-free likelihood.
Their nominal confidence threshold has not been calibrated for
finite-sample coverage.
These assumptions should be replaced by a detector-level
likelihood in a definitive reinterpretation.

The practical implication is that low-mass direct-detection
limits should be accompanied by an explicit halo-model band and
a clear description of the escape-speed and tail prescriptions.
For the models studied here, the SHM remains a useful reference
at high masses.
Near threshold, the velocity distribution is an essential part
of the reported result.

\section{Conclusions}
\label{sec:conclusions}

We have examined how Galactic dark matter halo modelling affects
the interpretation of LZ results.
Using twenty-two realizations spanning isotropic spherical
models, Osipkov--Merritt anisotropic models, axisymmetric Evans
models, and triaxial Gaussian distributions, we calculated
isotope-resolved recoil spectra and estimated the resulting
variation in published elastic-scattering limits.

For SI scattering, the halo-model dependence is modest above
approximately $40\,\GeV$: the limits remain within about
$\pm30\%$ of the isotropic reference.
Below approximately $15\,\GeV$, the envelope broadens to as much
as an order of magnitude because the signal depends
increasingly on the high-velocity tail.

Changing the escape speed from $504$ to $584\,\kms$ modifies the
high-mass limits by approximately $5\%$ or less, but can change
the $9\,\GeV$ limits by about a factor of two for classes A and B.
Larger variations arise when the imposed high-speed cutoff of
the class-C and class-D models is extended.

The SD results follow the same qualitative pattern.
Near $40\,\GeV$, however, nuclear-response uncertainty exceeds
the halo-model variation.
The published observed SD limits also lie near the lower edge
of the expected $1\sigma$ sensitivity band, which should be
retained when discussing experimental reach.

For an inelastic interpretation of the $248\,\keV$ event,
the high-velocity tail controls both the rate and the maximum
accessible mass splitting.
Some anisotropic models extend the allowed kinematic region
beyond the reference distributions, but this alone does not
establish consistency at the fixed Higgsino cross section.

\begin{acknowledgments}
While preparing this work, the authors used DeepSeekV4.1 to enhance readability and language. Following use of this tool, the authors reviewed and edited the content as necessary and accept full responsibility for the published content.
AA work was supported by the Program for Innovative Research Team in Anqing Normal University. 
Y.F.C. is supported in part by NSFC (12433002, W2611007), by CAS young interdisciplinary innovation team (JCTD-2022-20), by 111 Project (B23042), by USTC-PMO Introduction Fund for Science-Education Integration, by CSC Innovation Talent Funds, by USTC Fellowship for International Cooperation, and by USTC Research Funds of the Double First-Class Initiative.
\end{acknowledgments}

\bibliography{refs}

\end{document}